\documentclass[12pt]{article}
\def\lsim{\mathrel{\rlap{\lower4pt\hbox{\hskip1pt$\sim$}}
    \raise1pt\hbox{$<$}}}         
\def\gsim{\mathrel{\rlap{\lower4pt\hbox{\hskip1pt$\sim$}}
    \raise1pt\hbox{$>$}}}         

\usepackage{psfig}

\title{Supernova Neutrino-Nucleus Physics and the r-process}

\author{W. C. Haxton \\
Institute for Nuclear Theory and Department of Physics \\
Box 351550, University of Washington \\ 
Seattle, WA 98155\\ 
E-mail: haxton@phys.washington.edu}

\begin{document}
\begin{titlepage}
\maketitle

\begin{abstract}
This talk reviews three inputs important to neutrino-induced nucleosynthesis in a
supernova: 1) ``standard'' properties of the supernova neutrino flux, 
2) effects of phenomena like neutrino oscillations on that flux, 
and 3) nuclear structure issues in estimating cross sections for 
neutrino-nucleus interactions.  The resulting possibilities for 
neutrino-induced nucleosynthesis -- or the $\nu$-process -- in 
massive stars are discussed.  This includes two relatively recent extensions 
of $\nu$-process calculations to heavier nuclei, one focused
on understanding the origin of
$^{138}$La and $^{180}$Ta and the second on the effects following 
$r$-process freezeout.  From calculations of the neutrino post-processing of
the $r$-process distribution, 
limits can be placed on the neutrino fluence after freezeout and thus 
on the dynamic timescale for the expansion.
\end{abstract}
\end{titlepage}

\section{Basic Supernova Neutrino Characteristics}
A massive star, perhaps 15-25 solar masses, evolves through hydrostatic burning to 
an ``onion-skin'' structure, with a inert iron core produced from the explosive burning
of Si.  When that core reaches the Chandresekar mass, the star begins to collapse.
Gravitational work is done on the infalling matter, the temperature increases, and the
increased density and elevated electron chemical potential begin to favor 
weak-interaction conversion of protons to neutrons, with the emission of $\nu_e$s.  Neutrino
emission is the mechanism by which the star radiates energy and lepton number.   
Once the density exceeds $\sim$ 10$^{12}$ g/cm$^3$ in the infall of a Type II supernova,
however, neutrinos become trapped within the star by neutral-current scattering,
\begin{equation}
\tau_\nu^{diffusion} \gsim \tau^{collapse}
\end{equation}
That is, the time required for neutrinos to random walk out of the star exceeds $\tau^{collapse}$.  
Thus neither 
the remaining lepton number nor the gravitational energy released by further 
collapse can escape.

After core bounce a hot, puffy protoneutron star remains.  Over times on the 
order of a few seconds, much longer than the 100s of milliseconds required 
for collapse, the star 
gradually cools by emission of neutrinos of all flavors.  As the neutrinos 
diffuse outward, they tend to remain in flavor equilibrium through reactions 
such as
\begin{equation}
\nu_e + \bar{\nu}_e \leftrightarrow \nu_\mu + \bar{\nu}_\mu
\end{equation}
producing a rough equipartition of energy/flavor.  Near 
the trapping density of 10$^{12}$ g/cm$^3$ the neutrinos decouple, and this decoupling
depends on flavor due to the different neutrino-matter cross sections,
\begin{eqnarray}
\nu_x + e \leftrightarrow \nu_x + e:  \sigma_{\nu_\mu}/\sigma_{\nu_e} \sim 1/6 \nonumber \\
\nu_e + n \leftrightarrow p + e^+ ~~~~~~~~~~\nonumber \\
\bar{\nu}_e + p \leftrightarrow n + e^+.~~~~~~~~~ 
\end{eqnarray}
One concludes that heavy-flavor neutrinos, because of their weaker cross 
sections for scattering off electrons (and the absence of charged-current 
reactions off nucleons), will decouple at higher densities, deeper within 
the protoneutron star, where the temperature is higher.  In the case of electron 
neutrinos, the $\nu_e$s are more tightly coupled to the matter than 
the $\bar{\nu}_e$s, as the matter is neutron rich.  The result is the expectation 
of spectral differences among the flavors. If spectral peaks are used to define 
an effective Fermi-Dirac temperature, then supernova models \cite{supernova} 
typically yield 
values such as
\begin{eqnarray}
T_{\nu_\mu} \sim T_{\nu_\tau} \sim T_{\bar{\nu}_\mu} \sim T_{\bar{\nu}_\tau} \sim 8 MeV \nonumber \\
T_{\nu_e} \sim 3.5 MeV~~~~~T_{\bar{\nu}_e} \sim 4.5 MeV
\end{eqnarray}

Some of the issues relevant to subsequent neutrino-induced nucleosynthesis include: \\
$\bullet$ The $\nu_e$ and $\bar{\nu}_e$ temperatures are important for the p/n 
chemistry
of the ``hot bubble'' where the r-process is thought to occur.  
This is
high-entropy material near the mass-cut that is blown off the protoneutron star
by the neutrino wind. \\
$\bullet$ Matter-enhanced neutrino oscillations, in principle, could generate 
temperature inversions affecting p $\leftrightarrow$ n charge-current balance,
thus altering conditions in the ``hot bubble'' necessary for a successful $r$-process. \\
$\bullet$ If the ``hot bubble'' is the $r$-process site, then synthesized nuclei are
exposed to an intense neutrino fluence that could alter the r-process distribution. 
The relevant parameter is the neutrino fluence after r-process freezeout.

\section{New Neutrino Physics Discoveries and Potential Supernova Implications}
Following the chlorine, GALLEX/SAGE, and Kamioka/Super-Kamiokande experiments, 
strong
but circumstantial arguments led to the conclusion that the data indicated
new physics.  For
example, it was observed that, even with arbitrary adjustments 
in the undistorted fluxes of pp, $^7$Be, and $^8$B fluxes, the experimental results
were poorly reproduced \cite{Heeger}.  
When neutrino oscillations were included, however, several 
good fits to the data were found.  These included the small-mixing-angle (SMA) and
large-mixing-angle (LMA) MSW solutions, the LOW solution, and even the possibility
of ``just-so'' vacuum oscillations, where the oscillation length is comparable to
the earth-sun separation.  The ambiguities were convincingly removed by the
charged- and neutral-current results of SNO, which demonstrated that about 2/3rds
of the solar neutrino flux was carried by heavy-flavor neutrinos \cite{SNO}.

Similarly, anomalies in atmospheric neutrino measurements -- a zenith-angle
dependence in the ratio of electron-like to muon-like events -- indicated a 
distance-dependence in neutrino survival properties consistent
with oscillations.
The precise measurements of Super-Kamiokande provided convincing evidence 
for this
conclusion, and thus for massive neutrinos \cite{SK}.

A summary of recent discoveries in neutrino physics include: \\
$\bullet$ Oscillations in matter can be strongly enhanced. \\
$\bullet$ SNO identified a unique two-flavor solar neutrino solution corresponding
to $\theta_{12} \sim \pi/6$ and $\delta m_{12}^2 \sim 7 \times 10^{-5}$ eV$^2$. \\
$\bullet$ The KamLAND reactor $\bar{\nu}_e$ disappearance experiment has confirmed the
SNO conclusions and narrowed the uncertainty on $\delta m_{12}^2$ \cite{KamLAND}. \\
$\bullet$ The Super-Kamiokande atmospheric neutrino results show that those data
require a distinct $\delta m_{23}^2 \sim (2-3) \times 10^{-3}$ eV$^2$ and a mixing
angle $\theta_{23} \sim \pi/4$ that is maximal, to within errors. \\
$\bullet$ The KEK-to-Kamioka oscillation experiment K2K is consistent with the
Super-Kamiokande atmospheric results, finding $\delta m_{23}^2 \sim (1.5-3.9) \times
10^{-3}$ eV$^2$ under the assumption of maximal mixing \cite{K2K}.\\
$\bullet$ Chooz and Palo Verde searches for reactor $\bar{\nu}_e$ disappearance over
the $\delta m_{23}^2$ distance scale have provided null results, 
limiting $\theta_{13}$ \cite{Chooz}.

These results have determined two mass splittings, $\delta m_{12}^2$ and the
magnitude of $|\delta m_{23}^2|$.  But as only mass differences are known, the 
overall scale is undetermined.  Likewise, because the sign of $\delta m_{23}^2$ is
so far unconstrained, two mass hierarchies are possible: the ``ordinary'' one
where the nearly degenerate 1,2 mass eigenstates are light while eigenstate 3 is 
heavy, and the inverted case where the 1,2 mass eigenstates are heavy while
eigenstate 3 is light.  The relationship between the mass eigenstates
$(\nu_1,\nu_2,\nu_3)$ and the flavor eigenstates $(\nu_e,\nu_\mu,\nu_\tau)$ 
is given
by the mixing matrix, a product of the three rotations 1-2 (solar), 1-3,
and 2-3 (atmospheric):
\begin{equation}
\left( \begin{array}{c} \nu_e \\ \nu_\mu \\ \nu_\tau \end{array} \right)
= \left( \begin{array}{ccc} c_{12}c_{13} & s_{12}c_{13} & s_{13} e^{-i \delta} \\
-s_{12}c_{23}-c_{12}s_{23}s_{13} e^{i \delta} & c_{12}c_{23}-s_{12}s_{23}s_{13} e^{i \delta} & s_{23}c_{13} \\
s_{12}s_{23}-c_{12}c_{23}s_{13} e^{i \delta} & -c_{12}s_{23}-s_{12}c_{23}s_{13} e^{i \delta} & c_{23}c_{13} 
\end{array} \right) 
\left( \begin{array}{c} \nu_1 \\ \nu_2 \\ \nu_3 \end{array} \right) \nonumber
\end{equation}
\begin{equation}
= \left( \begin{array}{ccc} 1 & & \\ & c_{23} & s_{23} \\ & -s_{23} & c_{23} \end{array} \right)
\left( \begin{array}{ccc} c_{13} & & s_{13} e^{-i \delta} \\ & 1 & \\ -s_{13} e^{i \delta} & & c_{13} \end{array} \right)
\left( \begin{array}{ccc} c_{12} & s_{12} & \\ -s_{12} & c_{12} & \\ & & 1 \end{array} \right)
\left( \begin{array}{c} \nu_1 \\ e^{i \phi_1} \nu_2 \\ e^{i \phi_2} \nu_3 \end{array} \right) 
\end{equation}
Here $s_{12} = \sin{\theta_{12}}$, etc.
We see, in addition to the unknown third mixing angle $\theta_{13}$, this relationship
depends on one Dirac CP-violating phase parameterized by $\delta$ and two Majorana
CP-violating phases parameterized by $\phi_1$ and $\phi_2$.  The former could be 
measured in long-baseline neutrino oscillation experiments (with the ease of this
depending on the size of $s_{13}$), while the latter could influence rates for
double beta decay.

\begin{figure}[ht]
\begin{center}
\leavevmode
\psfig{bbllx=2.0cm,bblly=4.0cm,bburx=18.0cm,bbury=18.5cm,figure=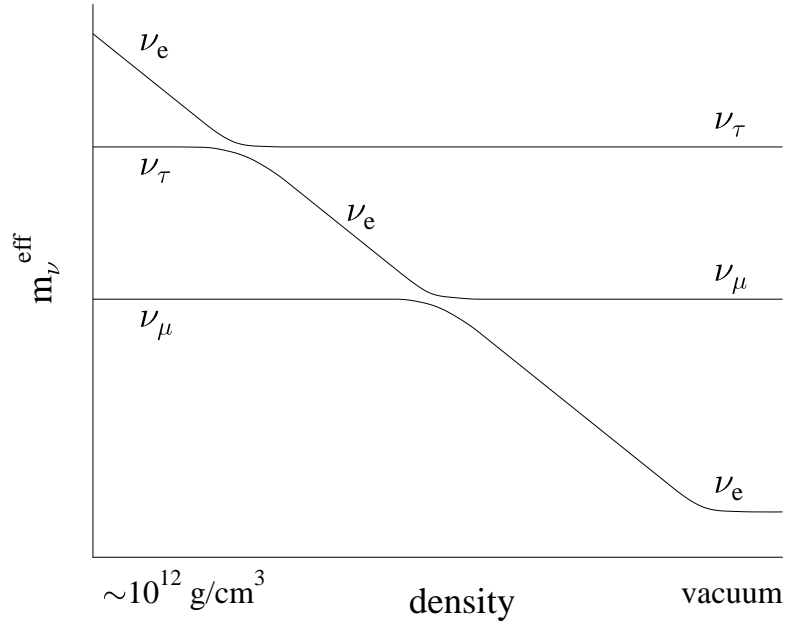,height=3.5in}  
\caption{A three-level MSW level-crossing diagram showing
the second crossing that should occur at densities characteristic of
the base of the carbon zone in a Type II supernova progenitor star.}
\end{center}
\end{figure}

These new neutrino physics discoveries could have a number of implications for
supernova physics: \\
$\bullet$ Because of solar neutrinos, we have been able to probe matter effects up
to densities $\rho \sim 100$ g/cm$^3$ characteristic of the solar core.  
As the density at the supernova neutrinosphere
is $\rho \sim 10^{12}$ g/cm$^3$, supernova neutrinos propagate
in an MSW potential that can be 10 orders of magnitude greater than any we
have tested experimentally.  
In addition, the neutrinos propagate in a dense neutrino 
background,
generating new MSW potential contributions due to $\nu-\nu$ scattering. Such
effects, as well as the magnitude of the ordinary-matter
MSW effects, may be unique to the supernova environment.\\
$\bullet$ We do not know $\theta_{13}$, which is the crucial mixing angle for
supernovae.  This angle governs the $\nu_e$-heavy flavor
level crossing encountered at depth in the star.  This crossing
occurs near the base of the carbon zone in the 
progenitor star, and remains adiabatic for $\sin^2{2 \theta_{13}} \gsim 10^{-4}$.
For the ordinary hierarchy, the resulting $\nu_e \leftrightarrow \nu_{\mu,\tau}$
crossing would lead to a hotter $\nu_e$ spectrum.  For an inverted hierarchy,
the crossing would be $\bar{\nu}_e \leftrightarrow \bar{\nu}_{\mu,\tau}$.\\
$\bullet$ Presumably the position of this crossing will be influenced by the
neutrino background contribution to the MSW potential.  This nonlinear problem
is rather complicated because the flavor content of the background evolves with
time (being $\nu_e$-dominated at early times).

\section{Nuclear Structure Issues}
Inelastic neutrino-nucleus interactions are important to a range of supernova 
problems,
including neutrino nucleosynthesis, the detection of supernova 
neutrinos in terrestrial
detectors, and neutrino-matter heating that could boost the 
explosion.  The heavy-flavor neutrinos have an average energy
$\langle E \rangle \sim 3.1 T \sim 25$ MeV.  However the most effective energy for
generating nuclear transitions can be substantially higher because 
cross sections grow
with energy and because nuclear thresholds are more easily overcome by neutrinos
on the high-energy tail of the thermal distribution.  

For the neutrino energy range of interest the allowed approximation, 
which includes only the Gamow-Teller $g_A \sigma(i) \tau_{\pm}(i)$ and 
Fermi $\tau_{\pm}(i)$ operators, is often
not adequate.  (These are given for charge-current reactions; the allowed operator
for inelastic neutral current reactions is $g_A \sigma(i) \tau_3(i)$.)
Important additional contributions come from operators that depend on the
three-momentum transfer $q$, which can approach twice the 
neutrino energy,
for back-angle scattering.  Consequently $qR$, where $R$ is the nuclear size, may 
not be small.  Such first-forbidden contributions may be as important 
as the allowed
contribution for supernova $\nu_\mu$s and $\nu_\tau$s.

For all but the lightest nuclei, cross sections must be estimated from nuclear
models, such as the shell model.  Shell model wave functions are generated 
by diagonalizing
an effective interaction in some finite Hilbert space of Slater determinants formed
from shells $|n(ls)jm_j\rangle$.  The space may be adequate for describing
the low-momentum components of the true wave function, but not the high-momentum
components induced by the rather singular short-range NN potential.  The effective 
interaction, usually determined empirically, is a low-momentum interaction 
that corrects
for the effects of the excluded, high-momentum excitations.  Similarly, effective
operators should be used in evaluating matrix elements, such as those of the weak
interaction operators under discussion here, and the shell model 
wave function
should have a nontrivial normalization.  In effective interaction theory, that
normalization is the overlap of the model-space wave function with the true wave
function.

Because effective interaction theory is difficult to execute properly -- in 
some sense
it is as difficult as solving the original problem in the full, infinite Hilbert
space -- nuclear modelers take short cuts.  Often all effective operator
corrections are
ignored: bare operators are used.  In other cases, phenomenological operator
corrections can been deduced from systematic comparisons of shell-model predictions
and experimental data.

The Gamow-Teller operator is an interesting case.  Rather thorough comparisons of
$2s1d$ and $2p1f$ shell-model predictions with measured allowed $\beta$-decay rates 
have yielded a simple, phenomenological effective operator:
the axial coupling $g_A^{eff} \sim 1.0$ should be used rather than 
the bare value \cite{Brown,Caurier}.
This observation is the basis for many shell-model estimates of the Gamow-Teller
response that governs allowed neutrino cross sections.  Many of the shell-model
techniques are quite powerful.  Moments techniques based on the 
Lanczos algorithm \cite{Caurier} 
have been used to treat spaces of dimension $\sim 10^8$: important supernova 
neutrino cross
sections for Fe and Ni isotopes have been derived in this way \cite{Langanke}.  
Another shell-model-based
method uses Monte Carlo sampling \cite{MonteCarlo}.

There are reasons to have less confidence in corresponding estimates of 
first-forbidden effective
operators.  The first-forbidden operators include the vector operator $q r(i)$ and
the axial-vector operators $[q \otimes r(i)]_{0,1,2}$.  Electron scattering and
photo-absorption provide tests of the vector operator, but direct probes of the
axial responses are lacking.  Unitarity is also an issue.  
Standard shell-model spaces satisfy the sum-rule contraints
for the Gamow-Teller operator: the operator cannot generate transitions outside 
a full shell, for example.  In contrast, for harmonic-oscillator Slater determinants,
the first-forbidden operators generate transitions for which $\Delta N = \pm 1$,
where $N$ is the principal quantum number.  Thus, underlying sum rules are
violated as the operators always connect either initial or final configurations
to states outside the shell-model space.

When the full momentum dependence of the weak interaction operators is included, the
resulting spin-spatial structure includes forms such as
\[ j_l (qr(i)) [Y_l (\Omega(i)) \otimes I_s]_{J M_J} \]
where $j_l$ is a spherical Bessel function, $Y_l$ a spherical harmonic, and $I_s$ is
a single-particle spin function.  (More complicated forms involve vector spherical
harmonics combined with spatial operators such as $\nabla(i)$.)  The fact that $q$ 
cannot then be factored from the operator then makes Lanczos moments techniques
less useful: at every desired $q$ the Lanczos procedure has to be repeated.
(There are techniques under development \cite{HNZ} which exploit special properties
of the harmonic oscillator to circumvent this problem.)  
Thus most calculations that treat the full momentum-dependence of the weak 
operators 
have used simple spaces, ones for which state-by-state summations of the weak
transition strengths are practical.  The approaches include truncated shell-model
spaces, models based on the Random Phase Approximation (RPA), and even the highly
schematic Goldhaber-Teller model.

Figure 2 compares continuum RPA results for charge-current reactions on
$^{16}$O with shell-model results of the sort described above \cite{Kolbe}.  The 
quantities plotted
are cross sections averaged over a thermal neutrino spectrum.  This is an 
interesting
test case because $^{16}$O, naively a closed-shell nucleus, has a smaller 
Gamow-Teller
response than most mid-shell nuclei.  Thus momentum-dependent contributions 
to the cross section should be more important than in many other cases.  
It is perhaps surprising,
given the assumptions implicit in both the shell-model and CRPA calculations,
that the results agree so well over the full range of interesting supernova neutrino
temperatures.  The only significant discrepancy, at very low temperatures, is due
to the inclusion of contributions from $^{18}$O in the shell model calculation
used in Fig. 2.
(The calculations were done for
a natural oxygen target.)  Due to the very low threshold for $^{18}$O
$\rightarrow$ $^{18}$F, this minor isotope 
(0.2\% relative to $^{16}$O)
dominates the O($\nu_e$,e) cross section at sufficiently low temperatures.     
The good agreement between the shell-model and CRPA calculations, of course, could
mask problems associated with common assumptions, such as the absence of a reliable
procedure for assessing effective operators beyond the allowed approximation.

\begin{figure}[ht]
\begin{center}
\leavevmode
\psfig{bbllx=-2.8cm,bblly=4.5cm,bburx=24.2cm,bbury=26.2cm,figure=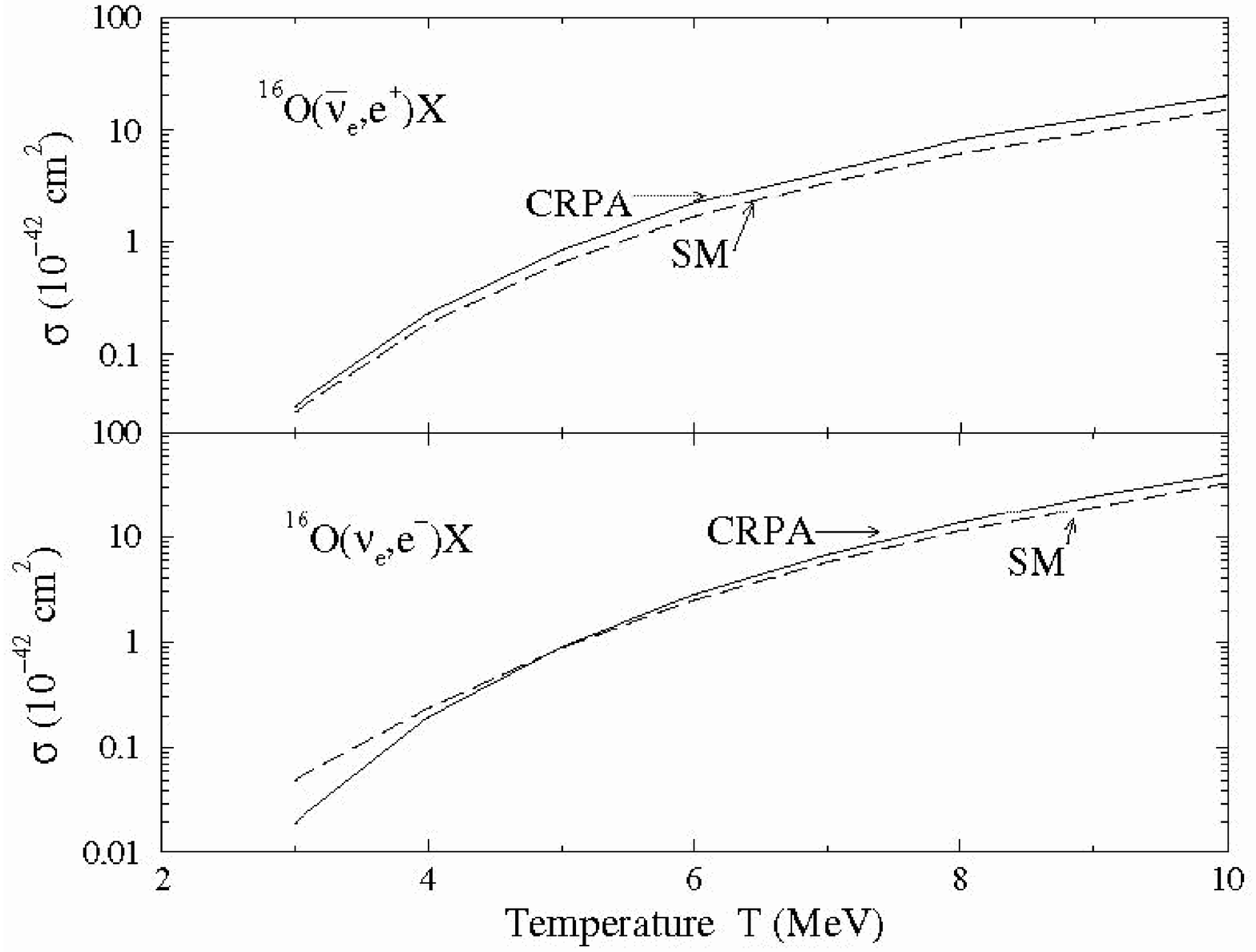,height=3.2in}   
\caption{Comparison of CRPA (full lines) and shell-model (dash lines) cross
section predictions, integrated over thermal neutrino spectra.  
The shell-model results include the contribution from $^{18}$O, important at
low temperatures in natural water for the $(\nu_e$,e$^-$) reaction.  From
Kolbe, Langanke, Martinez-Pinedo, and Vogel~\protect\cite{Kolbe}.}
\end{center}
\end{figure}

\section{The Neutrino Process}
Several rare isotopes are thought to be created during a core-collapse supernova
by neutrino reactions in the mantle of the star \cite{Woosley}.  The most common mechanism is
inelastic neutral-current neutrino scattering off a target nucleus like $^{20}$Ne
or $^{12}$C, with significant energy transfer, e.g., giant resonance excitation.
The nucleus, excited above the continuum, then
decays by nucleon or $\alpha$ emission, leading to new nuclei.  A nuclear network
calculation is required to assess the survival of the neutrino-process products,
such as $^{19}$F in the Ne shell and $^{11}$B in the C shell.  The co-produced 
nucleons can capture back on the daughter nucleus, destroying the product of 
interest.  Similarly, passage of the shock wave leads to heating that can
destroy the product by $(\gamma,\alpha)$ and similar reactions.  
Frequently the majority
of the instantaneous production is lost due to such explosive processing.

The enormous fluence of neutrinos can yield significant productions.  Typically
1\% of the nuclei in the deep mantle of the star -- the C, Ne, and O shells -- are
transmuted by neutrinos.  The most important products, like $^{19}$F and $^{11}$B,
tend to be relatively rare odd-A isotopes neighboring very plentiful parent nuclei,
such as $^{20}$Ne and $^{12}$C.  (The parent isotopes are the 
hydrostatic 
burning products, typically.)  The natural abundances of such odd-A isotopes
could, in principle, be due to neutrino processing.

Such nucleosynthesis calculations must be embedded in a model of the supernova
event.  Important factors include: \\
$\bullet$ a neutrino flux that tends to diminish exponentially, with a typical time
scale $\tau_\nu$ $\sim$ 3 sec; \\
$\bullet$ a pre-processing phase where nuclei in the mantle are exposed to the
neutrino flux at some fixed radius $r$, prior to shock arrival; \\
$\bullet$ a post-processing phase after shock wave arrival, where the material 
exposed to the neutrino flux is heated by the shock wave (potentially 
destroying
pre-processing productions), and then expands adiabatically off the star,
with a temperature $T$ that consequently declines exponentially; \\
$\bullet$ integration of these neutrino contributions into an explosive 
nucleosynthesis
network; and \\
$\bullet$ integration over a galactic model, with some assumptions on the range
of stellar masses that will undergo core collapse and mantle ejection.

Calculations of this nature were done by Woosley {\it et al.}
\cite{Woosley}.  Potentially
significant neutrino-process productions include the nuclei $^{19}$F, $^{10,11}$B,
$^7$Li, the gamma-ray sources $^{22}$Na and $^{26}$Al, $^{15}$N, $^{31}$P,
$^{35}$Cl, $^{39,40}$K, $^{51}$V, and $^{45}$Sc.  Although the nuclear reaction
network stopped at intermediate masses, the very rare isotopes $^{138}$La and
$^{180}$Ta were also identified as likely $\nu$-process candidates.

Recent work by Heger {\it et al.} \cite{Heger} extends 
these calculations in important ways.
First, the evolution of the progenitor star includes the effects of mass loss.
Second, a reaction network is employed that includes all of the heavy elements
through Bi, using updated reaction rates.  Third, the nuclear evaporation process --
emission of a proton, neutron, or $\alpha$ by the excited nucleus -- is treated
in a more sophisticated statistical model that takes into account known nuclear
levels and their spins and parities.  While the calculations lack a full set
of neutrino cross sections, those cross sections important to known ({\it e.g.},
$^{19}$F and $^{11}$B) and suspected ({\it e.g.}, $^{138}$La and $^{180}$Ta) 
neutrino products were
evaluated and incorporated into the network.

The results are shown in Fig. 3, with production factors normalized to that of
$^{16}$O.  Thus a production factor of one would mean that the $\nu$-process would fully
account for the observed abundance of that isotope.  While $^{11}$B might be
slightly overproduced and $^{19}$F slightly underproduced, given nuclear and
astrophysics uncertainties, the $\nu$-process yields of these
isotopes and $^{138}$La are compatible with this being their primary origin.
The case of $^{138}$La is particularly interesting, as the primary channel for
the production is charged-current reaction $^{138}$Ba($\nu$,e)$^{138}$La,
where $^{138}$Ba is enhanced in the progenitor star by the $s$-process.  This
production is the only known case where a charged-current channel dominates the
production.  Thus this yield is sensitive to the $\nu_e$ temperature -- a potential
indicator for oscillations if the transformation occurs deep within the star. 

\begin{figure}[ht] 
\begin{center}
\leavevmode
\psfig{bbllx=0.0cm,bblly=4.7cm,bburx=21.0cm,bbury=25.1cm,figure=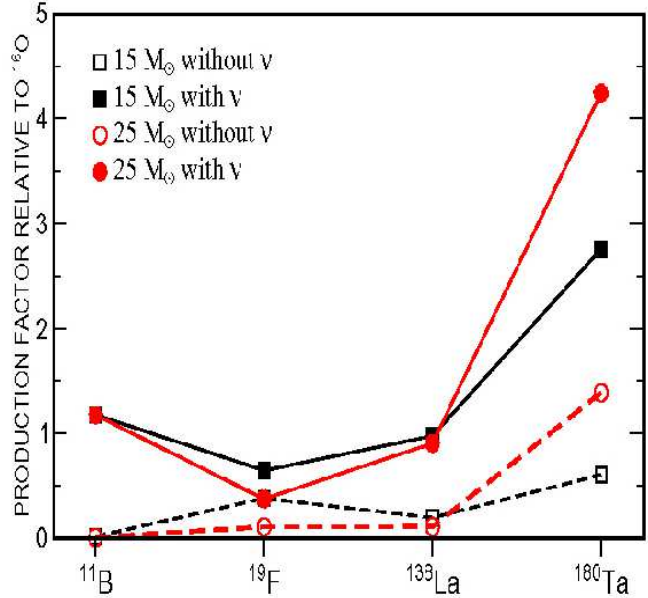,height=3.5in}   
\caption{Neutrino-process production factors for $^{11}$B, $^{19}$F, $^{138}$La,
and $^{180}$Ta, as calculated by Heger {\it et al.}~\protect\cite{Heger}.  The results are
normalized to the production of $^{16}$O in 15 $M_\odot$ (squares) and
25 $M_\odot$ (circles) progenitor stars.  The open (filled) symbols represent
stellar evolution studies in which neutrino reactions on nuclei were
excluded (included).}
\end{center}
\end{figure}

While $^{180}$Ta appears to be overproduced, the calculation does not distinguish
production in the 9$^-$ isomeric state from production in the 1$^+$ ground state.
Only the isomeric fraction should be counted.  An estimate of the $\nu$-process
fraction that ends up in the isomeric state, following a $\gamma$ cascade, has
not been made.  However, initiating reactions such as $^{181}$Hf($\nu_e$,e$^-$n)
involve low-spin parent isotopes, and the neutrino reaction transfers little angular
momentum.  Thus one would anticipate that the majority of the yield would
cascade to the ground state.  The reduction factor of 3-4 required to
bring the $^{180}$Ta production in line with the others of Fig. 3
is compatible with this.

There are other mechanisms for producing some of these $\nu$-process products.
One interesting one, for example, is cosmic-ray spallation reactions on CNO
nuclei in the interstellar medium, which can produce $^{10,11}$B and $^{6,7}$Li.
Some such process is required to explain the origin of $^{10}$B, for example.
Cosmic-ray production, a secondary process, and the $\nu$-process, a primary
mechanism, might be distinguished by measurements that would separately
determine the evolution of
$^{10}$B and $^{11}$B.  
If the $\nu$-process fraction of $^{11}$B
could be convincingly determined, this production would then become a more 
quantitative test of explosive conditions within the supernova carbon shell.

We note two recent observational results relevant to the $\nu$-process.
Prochaska, Howk, and Wolfe \cite{Prochaska} recently observed 
over 25 elements in a galaxy at 
redshift 
$z$ = 2.626, whose young age and high metallicity implies a nucleosynthetic
pattern dominated by short-lived, massive stars.  Their finding of a solar
B/O ratio in an approximately 1/3-solar-metallicity gas argues for a primary
(metal-independent) production mechanism for B such as the $\nu$-process, rather
than a secondary process.  Similarly, new F abundance data of Cunha {\it et al.}
show a low F/O ratio in two $\omega$ Centauri stars, which argues against AGB-star
production of F (one competing suggestion), but would be consistent with the 
$\nu$-process production \cite{Cunha}.

\section{Neutrino Process Effects in the $r$-process}
Other speakers have discussed the $r$-process and the likelihood
that the ``hot bubble'' -- the high-entropy nucleon gas that
is blown off the protoneutron star surface by the neutrino wind --
is the primary site for the r-process.
The nuclear physics of the $r$-process tells us that the synthesis
occurs when the neutron-rich nucleon soup 
is in the temperature range of 
$(3-1) \times 10^9$K, which, in the hot bubble $r$-process, might
correspond to a freeze-out radius of (600-1000) km and a time $\sim$ 10
seconds after core collapse.  The neutrino fluence after freeze-out
(when the temperature has dropped below 10$^9$K and the $r$-process
stops) is then $\sim$ $(0.045-0.015) \times 10^{51}$ ergs/(100km)$^2$.
Thus, after completion of the $r$-process, the newly synthesized
material experiences an intense flux of neutrinos.  This suggests
that $\nu$-process postprocessing could affect the
$r$-process distribution.

Comparing to our earlier discussion of carbon- and neon-zone
synthesis by the $\nu$-process, it is apparent that neutrino effects could
be much larger in the hot bubble $r$-process: the synthesis occurs {\it
much} closer to the star,
at $\sim$ 600-1000 km.  (The Ne-shell radius is $\sim$ 20,000 km.)
For this radius and a freezeout time of 10s, the
``post-processing" neutrino fluence -- the fluence
that can alter the nuclear distribution after the $r$-process is
completed -- is about 100 times larger than that responsible for
fluorine production in the Ne zone.  As approximately 1/300 of the
nuclei in the Ne zone interact with neutrinos, and noting that
the relevant neutrino-nucleus cross sections scale roughly as A (a
consequence of the sum rules that govern first-forbidden responses), one quickly
sees that the probability of a heavy $r$-process nucleus interacting with the
neutrino flux is approximately unity.

Because the hydrodynamic conditions of the $r$-process are highly
uncertain, one way to attack this problem is to work backward \cite{Qian}.
We know the final $r$-process distribution (what nature gives us) and we
can calculate neutrino-nucleus interactions relatively well.  Thus by subtracting
from the observed $r$-process distribution the neutrino
post-processing effects, we can determine what the $r$-process
distribution looked like at the point of freeze-out.  In Fig. 4,
the ``real" $r$-process distribution - that produced at freeze-out -
is given by the dashed lines, while the solid lines show the effects
of the neutrino post-processing for a particular choice of fluence.
The nuclear physics input into these calculations is precisely that
previously described: GT and first-forbidden cross sections, with the
responses centered at excitation energies consistent with those found
in ordinary, stable nuclei, taking into account the observed
dependence on $|N-Z|$.

\begin{figure}[htb] 
\begin{center}
\leavevmode
\psfig{bbllx=0.0cm,bblly=4.5cm,bburx=21.0cm,bbury=23.0cm,figure=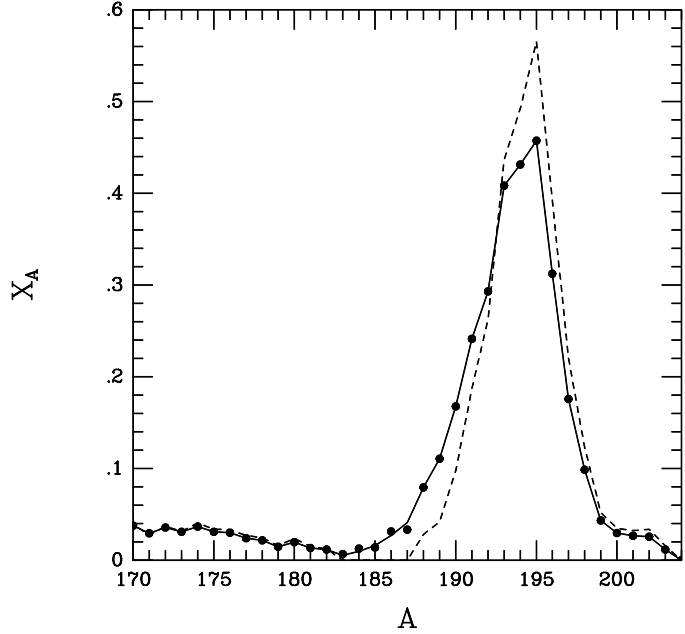,height=3.5in}   
\caption{Comparison of the $r$-process distribution that would 
  result from the freeze-out abundances near the A $\sim$ 195 mass peak
  (dashed line) to that where the effects of neutrino post-processing
  have been included (solid line).  The fluence has been fixed by
  assuming that the A = 183-187 abundances are entirely due to the
  $\nu$-process.}
\end{center}
\end{figure}
 
One important aspect of Fig. 4 is that the mass shift is
significant.  This has to do with the fact that a 20 MeV excitation of
a neutron-rich nucleus allows multiple neutrons ( $\sim$ 5) to be
emitted.  (The binding energy of the last
neutron in an $r$-process neutron-rich nuclei is about 2-3 MeV under
typical $r$-process conditions.)  The second thing to notice is that the
relative contribution of the neutrino process is particularly
important in the ``valleys" beneath the mass peaks: the reason is that
the parents on the mass peak are abundant, and the valley daughters
rare.  In fact, it follows from this that the neutrino process effects
can be dominant for precisely seven isotopes (Te, Re, etc.) lying in
the valleys below the A=130 (not shown) and A=195 (Fig. 4) mass peaks.
Furthermore if an appropriate neutrino fluence is
picked, these isotope abundances are correctly produced (within
abundance errors).  The fluences are
\begin{eqnarray}
     \mathrm{N} &=& 82~ \mathrm{peak}~~~~~0.031 \cdot 10^{51} 
\mathrm{ergs/(100km)^2/flavor} \nonumber \\
     \mathrm{N} &=& 126~ \mathrm{peak}~~~~0.015 \cdot 10^{51} 
\mathrm{ergs/(100km)^2/flavor}, \nonumber
\end{eqnarray}
values in fine agreement with those that would be found in a hot
bubble $r$-process.  So this is circumstantial but significant evidence
that the material near the mass cut of a Type II supernova is the site
of the $r$-process: there is a neutrino fingerprint.

A more conservative interpretation of these results, however, 
places a bound on the $r$-process post-processing neutrino
fluence by insisting that these isotopes not be overproduced.  This bound
will hold even if there are other mechanisms, such as neutron emission accompanying 
the $\beta$ decay of $r$-process parent nuclei as they move to the valley
of stability, that contribute to the abundances of these rare isotopes.
This bound is then an interesting constraint on supernova dynamics:
the neutrino fluence after freezeout depends on the flux at the 
time of freezeout 
and on the dynamic time scale (or the velocity of the material being
expelled from the supernova). This 
constraint is plotted in Fig. 5 along with one imposed by
the observed $\beta$-flow equilibrium of nuclei near the mass peak.
(The $\beta$-flow equilibrium requires that the neutrino flux at 
freezeout not exceed the value where the neutrino reactions would 
compete with $\beta$ decay.  This would destroy the observed correlation
between abundance and $\beta$ decay lifetime.) Together these two 
constraints place upper bounds on the luminosity at freezeout (equivalently,
a lower bound on the freezeout radius) and on the dynamic timescale.

\begin{figure}[ht] 
\begin{center}
\leavevmode
\psfig{bbllx=0.0cm,bblly=5.0cm,bburx=19.8cm,bbury=23.0cm,figure=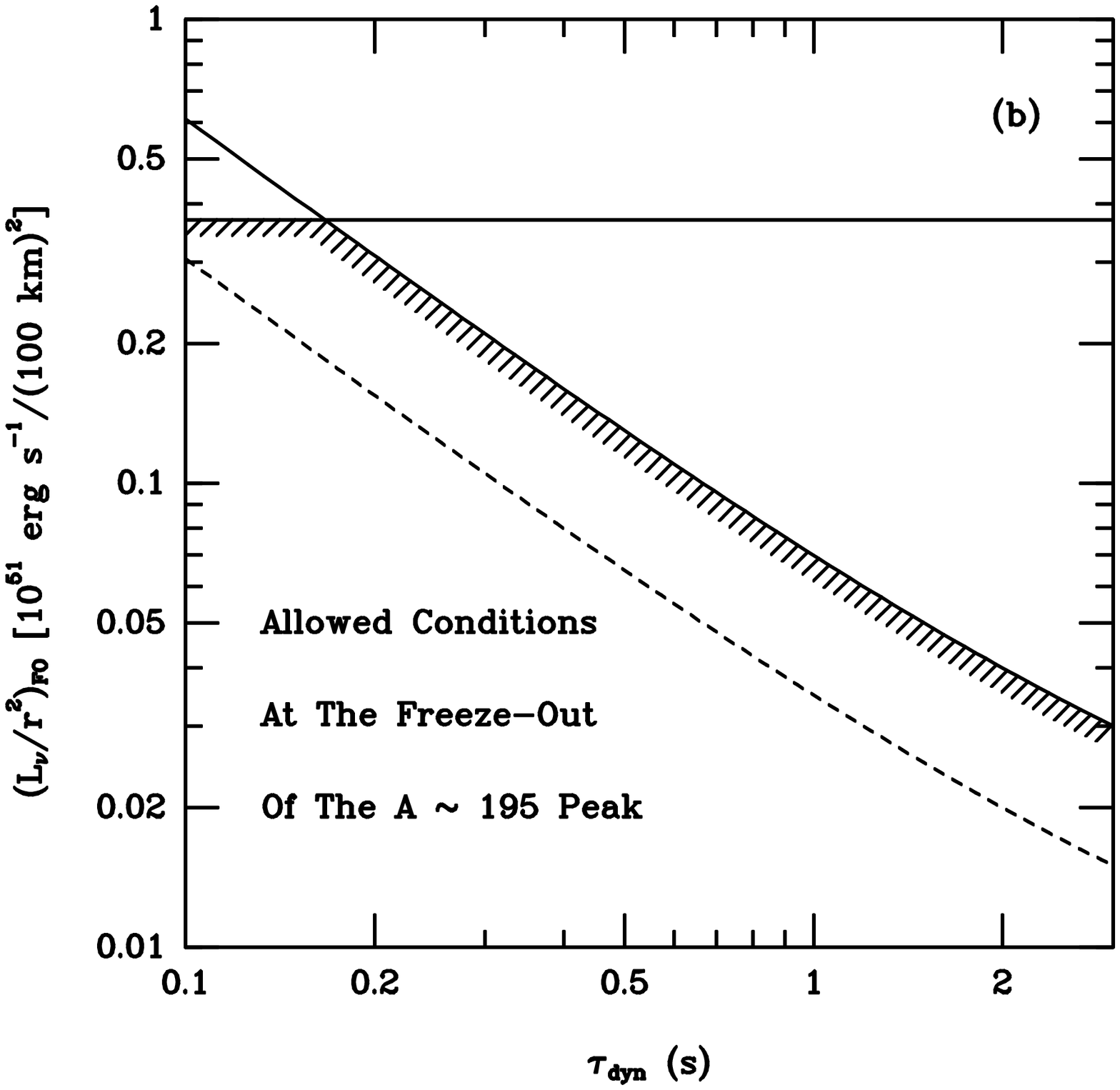,height=3.5in}   
\caption{Constraints imposed on the neutrino flux parameter $L_\nu/r^2$ at
freezeout, where $L_\nu$ is the luminosity and $r$ the freezout radius,
and on the dynamic timescale $\tau_{dyn}$ governing the expansion.
The horizontal solid line results from the condition of $\beta$-flow
equilibrium for the A=195 peak.  The diagonal solid line is the
requirement that $\nu$-process post-processing of the $r$-process peak
not overproduce nuclei in the mass region A=183-187.  The calculation
assumes a neutrino flux that evolves exponentially with $\tau_\nu$ = 3s.
Parameters lying on the dashed line corresponds to the fluence determined
by attributing the full abundances of A=183-187 nuclei to the neutrino
process.}
\end{center}
\end{figure}

\section{Summary}
This goal of this talk is to make some connections between supernova
neutrino physics, the nuclear structure governing neutrino-nucleus interactions,
and new neutrino properties.  The main example used here, the neutrino process,
connects observable abundances with supernova properties, such as the $r$-process 
freezeout
radius and dynamic timescale.  Thus by identifying $\nu$-process products and
by reducing the associated nuclear structure uncertainties that govern their
abundances, one may be
able to place significant constraints on the explosion mechanism.

The conditions for the $r$-process itself and
for various $\nu$-process productions are set by neutrino physics.
For example, the p/n chemistry of the ``hot bubble'' is largely governed by
charged-current p $\leftrightarrow$ n reactions, 
while the productions of $^{138}$La and $^{19}$F
depend primarily on charged-current interactions on $^{138}$Ba
and on neutral-current reactions on $^{20}$Ne, respectively.
Thus in principle such productions could be influenced by oscillations that
invert $\nu_e$ and heavy-flavor neutrino spectra (or, in the case of an inverted
hierarchy, $\bar{\nu}_e$ and heavy-flavor antineutrino spectra).  This is another
important reason for exploring the nucleosynthetic ``fingerprints'' of 
supernova neutrinos.  The productions identified so far that would be influenced
by neutrino oscillations, such as $^{138}$La, are created at densities above
those characterizing the naive 1-3 MSW matter crossing, 
at least in the pre-processing
phase.  However, we have noted that the 
location of MSW crossings could be perturbed
by neutrino background effects.  Furthermore, in the post-processing phase,
crossings will arise in the rarified matter that expands off the star.  This is
a fascinating question for the $r$-process, and perhaps also for certain 
$\nu$-process productions.  These observations should motivated further studies
of the potential effects of new neutrino physics on supernova nucleosynthesis.

This work was supported in part by the Office of Science, U.S. Department of
Energy, under grants DE-FG02-00ER41132 and DE-FC02-01ER41187.

\end{document}